\documentclass[twocolumn,superscriptaddress,showkeys,prb]{revtex4}
\usepackage{mathrsfs}
\usepackage{amsfonts}
\usepackage[mathscr]{eucal}
\usepackage{amssymb}
\usepackage{amsmath}
\usepackage{dcolumn}
\usepackage{bm}
\usepackage{xcolor}
\usepackage{graphicx}
\begin{document}

\newcommand\ppcf{Plasma Phys. Control. Fusion }
\newcommand\pop{Phys. Plasmas }

\title{Geodesic Acoustic Mode in Toroidally Rotating Anisotropic Tokamaks}
\author{Haijun Ren}
\email{hjren@ustc.edu.cn}
\affiliation{The Collaborative Innovation Center for Advanced Fusion Energy and Plasma Science, and Department of Modern Physics, University of Science and Technology of China, Hefei 230026, P. R. China}


\begin{abstract}
Effects of anisotropy on the geodesic acoustic mode (GAM) is analyzed by using gyro-kinetic equations applicable to low-frequency microinstabilities in a toroidally rotating tokamak plasma. Dispersion relation in the presence of arbitrary Mach number $M$, anisotropy strength $\sigma$, and the temperature ration $\tau$ is analytically derived. It is shown that when $\sigma$ is less than $ 3 + 2 \tau$, the increased electron temperature with fixed ion parallel temperature increases the normalized GAM frequency. When $\sigma$ is larger than $ 3 + 2 \tau$, the increasing of electron temperature decreases the GAM frequency. The anisotropy $\sigma$ always tends to enlarge the GAM frequency. The Landau damping rate is dramatically decreased by the increasing $\tau$ or $\sigma$.
\end{abstract}


\date{\today}

\maketitle

\section{Introduction}
Tokamak plasmas are usually described by isotropic equilibria, even in the low-collisional or collisionless limit\cite{book} . It is a reasonable approximation for Ohmically heated discharges but becomes quite unrealistic in the presence of intense auxiliary heating. It is known that the neutral beam injection (NBI) heating, ion cyclotron resonance heating (ICRH), and electron cyclotron resonance heating (ECRH) can produce strong plasma anisotropy\cite{Iter}. More specifically, NBI generally creates anisotropic equilibria with $p_\parallel > p_\perp$, while the ICRH typically drives perpendicular dominant anisotropy with $p_\perp > p_\parallel$, where $p_\parallel (p_\perp)$ is the parallel (perpendicular) pressure with respect to the magnetic field\cite{Iter}. In some experiments at JET\cite{Jet1}, an ICRH-heated minority ion population with strong temperature anisotropy was observed. In MAST, the beam pressure ratio $p_\perp/p_\parallel \simeq 1.7$ was achieved during the NBI heating\cite{Mast}. Plasma equilibrium with strong anisotropic effects was also found in the low density discharges of the Large Helical Device (LHD) with powerful tangential NBI\cite{LHD}.

The theory of tokamak anisotropic equilibria and stability has been approached long ago\cite{1958,1964,1967,1970,1974,1974b,1978,1980,1981} by many authors\cite{1987,1990,2010,2012,2012b,2013,2014} devoted to the equilibria by using magnetohydrodynamic (MHD) model or the guiding-center theory (see, for example, Refs. \onlinecite{1964,1970,1990,2013}), and to the stability by using energy principle\cite{1974,1976,1995}. The importance of the pressure anisotropy effect on the plasma equilibrium in tokamaks was emphasized in Ref. \onlinecite{Jet2} by implementing anisotropic pressure model into the equilibrium reconstruction code EFIT and applying the algorithm to selected high-performance discharges of the tokamaks JET and Tore Supra. It was shown that due to the assumption of isotropic pressure, earlier attempts using the pressure from transport analysis codes often gave unrealistic results\cite{LHD,Jet2}. While the anisotropic pressure model can yield a consistent equilibrium, showing a large difference in parallel and perpendicular pressures during additional heating\cite{2003}. One basic outcome of anisotropic equilibria is that the outward centrifugal shifts of the magnetic axis and of the mass density are proved to be increased by the anisotropy with $p_\perp > p_\parallel$ or decreased by $p_\perp < p_\parallel$\cite{1990,2004}. The other is that the two pressures $p_\perp$ and $p_\parallel$ and plasma density $\rho$ are no longer flux functions and depend on the poloidal angle through the magnetic field strength $B$. The importance of the poloidal asymmetry in the dispersion relation of geodesic acoustic mode (GAM) was checked in Ref. \onlinecite{HR2014}, which is limited to the case of zero electron temperature. However, in normal conditions, there is $T_e/T_i \gg 1$ in the core and $T_e/T_i \thicksim 1$ at the edge of the plasma column, where $T_e$ and $T_i$ are the electron and ion temperatures, respectively. It is of great importance to take into account electron temperature effect. Besides, considering that plasma rotation is also driven by auxiliary heating so that a rigid anisotropic equilibrium should contain the rotation, which motivated the present work devoted to the kinetic GAM in an isotropic toroidally rotating tokamak.

GAM is basically a sort of electrostatic perturbation with toroidally symmetrical and poloidally nearly symmetrical structure\cite{Diamond}. It has attracted intense attention since theoretically predicted by Winsor \emph{et al.} in 1968\cite{winsor}. The physical interpretation of GAM is that the radial drift of the perturbed particles distribution due to the geodesic curvature effect compensates the polarization drift and an oscillation with a frequency around the ion transit frequency is excited. By using drift kinetic equation, the GAM in the plasma with bi-Maxwellian distribution for ions was investigated and the previous kinetic result was recovered when zeroing the anisotropy\cite{Zhou2010}. However, the equilibrium distribution function used in Ref. \onlinecite{Zhou2010} actually violated the MHD equilibrium condition. In the present work, we will derive the dispersion relation of GAM in an anisotropic tokamak plasma by using the modified gyro-kinetic (GK) equation applicable to low-frequency microinstabilities in a rotating axisymmetric plasma and check the effects of ion anisotropy and the ratio between electron and ion temperatures on the GAM frequency and Landau damping rate.

\section{Equilibrium}

Let us consider a large-aspect-ratio tokamak plasma with a toroidally symmetric magnetic field $\vec{B} = I (\psi) \nabla \zeta + \nabla \zeta \times \nabla \psi$, and work in the $(r, \theta, \zeta)$ coordinate system, where $\psi(r)$ is the magnetic flux, and $\zeta$ and $\theta$ are the toroidal and poloidal angles, respectively. Circular cross-section is assumed in our calculation with $R = R_0 + r \cos \theta$. The subscript $0$ denotes the equilibrium profile and the prefix $\delta$ denotes the perturbed one, but the subscript is omitted and the equilibrium magnetic field is referred to by $\vec{B}$ directly. Only toroidal rotation is taken into account with $\vec{u}_0 = \omega_T(\psi) R^2 \nabla \zeta$. $\vec{w} = \vec{v} - \vec{u}_0$ is the particle velocity in the local reference frame moving with $\vec{u}_0$ relative to the lab frame. In the local frame, a standard anisotropic Maxwellian distribution function with two temperatures is
\begin{align}
\label{eq1}
F_0^j = n_0^j(\psi, \theta)  \frac{(2 \pi/m_j)^{ - 3/2}}{T_\parallel^{1/2} T_\perp} e^{ - \frac{m_j w_\parallel^2}{2 T_\parallel} - \frac{m_j w_\perp^2}{2 T_\perp}},
\end{align}
where $n_0^j$ is the number density and $m_j$ is the mass of species $j$ ($j = i, e$ for ions and electrons, respectively). The zeroth-order Fokker-Planck equation reads\cite{Hinton1985}
\begin{align}
w_\parallel & \vec{b} \cdot \nabla F_0 - \bigg[ \frac{q_j}{m_j} \vec{b} \cdot \nabla \Phi_0 + (\vec{u}_0 \cdot \nabla ) \vec{u}_0 \cdot \vec{b} \bigg] \frac{\partial f_0}{\partial w_\parallel} \nonumber\\
 & + \frac{1}{2} w_\perp^2 (\nabla \cdot \vec{b}) \bigg(\frac{\partial F_0}{\partial w_\parallel} - \frac{w_\parallel}{w_\perp} \frac{\partial F_0}{\partial w_\perp} \bigg) = 0,
\end{align}
in which $q_j(= \pm e)$ is the charge of species $j$. After substitution of \eqref{eq1}, the equation above contains terms with three types of velocity dependence, proportional to $w_\parallel$, $w_\parallel^3$, and $w_\parallel w_\perp^2$. In order that the equation be true for all $w$, it is necessary that the coefficients of each of these three types of terms vanish separately:
\begin{align}
\label{motion}
&\vec{b} \cdot ( \nabla \ln n_0^j  - \nabla \ln T_\perp ) + \frac{1}{T_\parallel} \vec{b} \cdot \nabla (q_j \Phi_0 + m_j \vec{u}_0 \cdot \nabla \vec{u}_0) = 0,\\
\label{tpara}
& \vec{b} \cdot \nabla T_\parallel = 0,\\
\label{tperp}
& \vec{b} \cdot \nabla \ln T_\perp + \bigg(1 - \frac{T_\perp}{T_\parallel} \bigg) (\nabla \cdot \vec{b} ) = 0.
\end{align}
Eq. \eqref{tpara} indicates $T_\parallel = T_\parallel (\psi)$, implies that parallel temperature gradient is forbidden in the local thermal equilibrium state\cite{Hinton1985}. According to Eq. \eqref{tperp}, the perpendicular temperature has the following form
\begin{align}
\label{perp}
T_\perp (\psi, B ) = \frac{B T_\parallel(\psi)}{B - B_0(\psi)}.
\end{align}
Meanwhile, electrons are supposed to have a standard Maxwellian distribution function with temperature $T_e$. It is reasonable since the temperature relaxation rate between electron perpendicular and parallel temperatures is the greatest\cite{Chao} and the time scale for electron relaxation can be much shorter than the ion time scale\cite{Hinton1985}, which requires $\tau \gg m_e/m_i$. As a result, the ion version of Eq. \eqref{motion} yields the ions equilibrium density as
\begin{align}
n_0^i = \frac{T_\perp}{T_\parallel} N(\psi) e^{- \frac{e \widetilde{\Phi_0}}{T_\parallel} + \frac{m_i \omega_T^2 \widetilde{R^2}}{2 T_\parallel}}.
\end{align}
As a naturally result of $\tau \gg m_e/m_i$, the mass in the electron version of Eq. \eqref{motion} is negligible. One then obtains the electron number density
\begin{align}
n_0^e = N_e(\psi) e^{\frac{e \widetilde{\Phi_0}}{T_e}}.
\end{align}
Here, we write $\Phi_0 = \langle \Phi_0 \rangle + \widetilde{\Phi_0}$, where $\langle \Phi_0 \rangle$ is the magnetic surface average and $\widetilde{\Phi_0}$ is the part depending on the poloidal angle. Using charge neutrality $n_0^e = n_0^i = n_0$ and $\langle n_0^i \rangle = \langle n_0^e \rangle$, we arrives at
\begin{align}
\label{phi}
e \Phi_0 = \frac{T_e}{1 + \tau} \ln \bigg(\frac{T_\perp}{\langle T_\perp \rangle} \bigg)+ \frac{T_e}{1 + \tau} M^2,
\end{align}
where $\tau$ stands for the temperature ratio $T_e/T_\parallel$ and $M$ is the Mach number defined as $\omega_T R/v_{T \parallel}$ with the ion parallel thermal velocity $v_{T \parallel} = \sqrt{2 T_\parallel/m_i}$.

\section{Dispersion Relation}

The perturbed distribution function is determined by the modified GK equation applicable to low-frequency microinstabilities in a toroidally rotating tokamak, which shows\cite{GK,book,CGK1968}
\begin{align}
\delta F_j = q_j \frac{\partial F_0^j}{\partial E}\delta \phi + [1 - J_0^2(k_r \rho_j) ] \frac{q_j}{B} \frac{\partial F_0^j}{\partial \mu} \delta \phi + J_0 \delta h_j,
\end{align}
in which $\delta h_j$  is governed by
\begin{align}
& \bigg[\frac{\partial }{\partial t} +  (w_\parallel \vec{b} + \vec{u}_0 + \vec{v}_D ) \cdot \nabla \bigg] \delta h_j\nonumber\\
= & - q_j J_0 \frac{\partial F_0^j}{\partial E} \bigg( \frac{\partial }{\partial t} + \vec{u}_0 \cdot \nabla \bigg) \delta \phi - J_0 \frac{q_j}{m_j \omega_c^j} \vec{b} \times \nabla \delta \phi \cdot \nabla F_0^j \nonumber\\
+ & J_0 \frac{q_j}{\omega_c^j} \vec{b} \times \nabla \delta \phi \cdot [(w_\parallel \vec{b} + \vec{u}_0) \cdot \nabla \vec{u}_0 + \nabla \vec{u}_0 \cdot (w_\parallel \vec{b} + \vec{u}_0)] \frac{\partial F_0^j}{\partial E}.
\end{align}
Only perturbed electrostatic potential is taken into account, which is justified for electrostatic GAM in a low-$\beta$ plasma. Here, $k_r$ is the radial wave number of GAM, $\delta \phi$ is the perturbed electrostatic potential, $J_0$ is the zeroth-order Bessel function, $\rho_j = w_\perp/\omega_c^j$ is the Larmor radius, $\omega_c^j = q_j B/m_j$ is the gyro frequency, and $\vec{v}_D$ is the leading order drift velocity. In the lab frame, electric field can be expanded as $\vec{E} = \vec{E}_{- 1} + \vec{E}_0 + \cdots$, where $\vec{E}_{- 1} = - \vec{u}_0 \times \vec{B}$ and $\vec{E}_0$ is related to $\Phi_0$, displayed in Eq. \eqref{phi}. Hence in the local reference frame, particles only feel the potential $\Phi_0$. As a result, the drift velocity can be expressed as\cite{GK}
\begin{align}
&\vec{v}_D = [(w_\parallel^2 + w_\perp^2/2)/\omega_c^j] \vec{b} \times \nabla \ln B\nonumber\\
&+ \frac{\vec{b}}{\omega_c^j} \times \bigg[\frac{q_j}{m_j} \nabla \Phi_0 + \vec{u}_0 \cdot \nabla \vec{u}_0 + w_\parallel (\vec{b} \cdot \nabla \vec{u}_0 + \vec{u}_0 \cdot \nabla \vec{b}) \bigg].
\end{align}
Meanwhile, the magnetic moment $\mu = \frac{1}{2 B} m_j w_\perp^2$ and the energy $E$ is defined as
\begin{align}
E = \frac{1}{2} m_j w^2 - \frac{1}{2} m_j u_0^2 + q_j \Phi_0.
\end{align}
The ion equilibrium distribution \eqref{eq1} can be rearranged as
\begin{align}
\label{fenbu}
F_0^i = \overline{N}(\psi) (\pi v_{T \parallel}^2)^{- \frac{3}{2}} e^{ - \frac{E}{T_\parallel} + \frac{\mu B_0}{T_\parallel}}.
\end{align}

We focus on the ions perturbed distribution function first. Using the properties of $\vec{u}_0$, one can show that $\nabla \vec{u}_0 = \omega_T R (\nabla R \nabla \zeta - \nabla \zeta \nabla R) + R^2 \nabla \omega_T \nabla \zeta$ and $\vec{u}_0 \cdot \nabla \vec{b} = \vec{b} \cdot \nabla \vec{u}_0$. Recalling the expression of $\Phi_0$ in \eqref{phi}, we have
\begin{align}
\vec{v}_D \cdot \nabla \psi = & \,\frac{I B}{\omega_c^i} (w_\parallel^2 + w_\perp^2/2) \nabla_\parallel \bigg( \frac{1}{B} \bigg) + \frac{\omega_T^2 I}{2 \omega_c^i (1 + \tau)} \nabla_\parallel R^2\nonumber\\
&  + \frac{w_\parallel \omega_T B}{\omega_c^i} \nabla_\parallel R^2 + \frac{I \tau T_\parallel}{e (1 + \tau)} (1 - \sigma) \nabla_\parallel \bigg( \frac{1}{B} \bigg),
\end{align}
where $\sigma$ is short for $T_\perp/T_\parallel$, representing the anisotropy strength. It is also found
\begin{align}
(w_\parallel \vec{b} & + \vec{u}_0) \cdot \nabla \vec{u}_0 + \nabla \vec{u}_0 \cdot (w_\parallel \vec{b} + \vec{u}_0) \nonumber\\
&  = \bigg( \frac{I w_\parallel}{B} + \omega_T R^2 \bigg) \nabla \omega_T.
\end{align}
In view of the two equations above, the GK equation is reduced to
\begin{align}
\partial_\theta \delta h_i - i n_d^i \sin \theta \delta h_i - i \frac{\omega}{\omega_t^i} \delta h_i = i e J_0 \frac{\omega}{\omega_t^i} \frac{\partial F_0^i}{\partial E} \delta \phi.
\end{align}
Here, $\omega_t^i = w_\parallel/(q R)$ is the modified transit frequency and $n_d^i = k_r \delta_b^i$, where $\delta_b^i$ is the ions orbit width defined as $\frac{1}{\omega_t^i \omega_c^i R} \bigg[ w_\parallel^2 + \frac{1}{2} w_\perp^2 + \frac{M^2 v_{T \parallel}^2}{1 + \tau} + 2 w_\parallel v_{T \parallel} M  + \frac{\tau v_{T \parallel}^2 }{2 (1 + \tau)} (1 - \sigma) \bigg]$. The disturbed distribution function can be easily solved as
\begin{align}
\label{ionp}
 \delta F_i  & =  [1 - J_0^2 (k_r \rho_i)] \bigg( \frac{\partial F_0^i}{\partial E} + \frac{1}{B} \frac{\partial F_0^i}{\partial \mu} \bigg) e \delta \phi + e J_0^2  \frac{\partial F_0^i}{\partial E}\nonumber\\
& \times \sum_{n, k} i^{n - k} J_{n + l - k} (n_d^i) J_l(n_d^i) \frac{(l - k) \delta \phi_n e^{i k \theta}}{l - k + \omega/\omega_t^i} .
\end{align}
In view of \eqref{perp} and \eqref{fenbu}, one has $\frac{\partial F_0^i}{\partial E} + \frac{1}{B} \frac{\partial F_0^i}{\partial \mu} = - \frac{1}{T_\perp} F_0^i$.

Generally for ions, the finite-Larmor-radius (FLR) effect is taken into account by assuming $k_r \rho_i \thicksim \Delta \ll 1$ and the finite-orbit-width (FOW) effect is also considered by assuming $n_d^i \thicksim \Delta$. While for electrons, we have $k_r \rho_e \simeq 0$ and $n_d^e \simeq 0$. The electron disturbed distribution function is simplified to
\begin{align}
\label{elecp}
\delta F_e = \frac{e}{T_e} F_0^e \sum_{k \neq 0} \delta \phi_k e^{i k \theta}.
\end{align}

The dispersion relation of GAM is derived by using quasi-neutrality condition, $\delta n_i = \delta n_e$, namely, $\int d^3 w \delta F_i = \int d^3 w \delta F_e$. Inserting $\delta F_j $ into the quasi-neutrality condition gives birth to the dispersion relation. Before doing that, we need to simplify $\delta F_i$ by cutting off the coupling chains. Considering $\delta \phi_n \thicksim (k_r \rho^i)^n \delta \phi_0$, we take into account only the coupling between $\delta \phi_0 $ and $\delta \phi_{\pm 1}$ by neglecting all high-order harmonics. Here, $\rho^i = v_{T \parallel}/\omega_c^i$ is the ion Larmor radius. Consequently, we write $\delta F_i = \delta F_i^0 + \delta F_i^{\pm 1} e^{\pm i \theta}$. Eventually, using Eqs. \eqref{ionp} and \eqref{elecp} and the quasi-neutrality condition, $\delta \phi_k$ is obtained as $[1 + \tau (1 + \zeta \mathcal{Z}(\zeta) ) ] \delta \phi_k = - i \frac{1}{2} q k_r \rho^i \tau k \mathcal{G}(k M) \delta \phi_0$ for $k = \pm 1$. For $k = 0$, we find
\begin{align}
\mathcal{S}(M) \delta \phi_0 - \frac{i}{q k_r \rho^i} \mathcal{G}(M) \delta \phi_1 + \frac{i}{q k_r \rho^i} \mathcal{G} (- M) \delta \phi_{- 1} = 0.
\end{align}
Finally, we obtain the following dispersion relation
\begin{align}
\label{dis}
\mathcal{S} - \frac{1}{2} \tau \frac{\mathcal{G}^2(M) + \mathcal{G}^2(- M)}{ 1 + \tau (1 + \zeta \mathcal{Z}(\zeta))} = 0,
\end{align}
in which we have denoted
\begin{align}
\mathcal{S} & = \,\frac{1}{q^2} + \frac{1}{2} + \frac{\tau + \sigma}{1 + \tau} + \frac{6 + 4 \tau}{1 + \tau} M^2 + \zeta^2 + \frac{\mathcal{Z}(\zeta)}{\zeta} \bigg[ \zeta^4 \nonumber\\
 + & \bigg(\frac{\tau + \sigma}{1 + \tau} + \frac{6 + 4 \tau}{1 + \tau} M^2 \bigg) \zeta^2 + \frac{M^2}{(1 + \tau)^2} (\sigma + \tau + M^2 ) \nonumber\\
& + \frac{\sigma (\tau + \sigma)}{2 (1 + \tau)} + \frac{\tau^2 (1 - \sigma)^2}{4 (1 + \tau)^2} \bigg],\\
\label{G}
\mathcal{G}(M) & =  \,\zeta + 2 M + \bigg(\zeta^2 + 2 \zeta M + \frac{\tau + \sigma}{2 (1 + \tau)} + \frac{M^2}{1 + \tau} \bigg) \mathcal{Z}(\zeta).
\end{align}
Here, $\zeta$ is the normalized frequency defined as $ q \omega R/v_{T \parallel}$ and $\mathcal{Z} (\zeta) $ is the plasma dispersion function. Letting $M = 0$ and $\sigma = 1$ in the equation above reproduces the result in Ref. \onlinecite{Gao2008}. For just $\sigma = 1$, the dispersion relation of GAM in a toroidally rotating isotropic plasma is recovered\cite{HR2015}. Assuming $M = 0$ and $\tau = 0$ in Eq. \eqref{dis} yields the result identical with the one in Ref. \onlinecite{HR2014}.

\section{Discussion and Conclusion}
Explicit analytical solutions to the dispersion relation \eqref{dis} for arbitrary $\zeta$ are difficult to obtain. Here we are restricted to the GAM with $\zeta \gg 1$ in order to find an asymptotic solution, which requires large safety factor. Hence, analytical results below are expected to be accurate enough only when $q$ is sufficiently high, i.e., $q \geqslant 3$ as shown in some previous literature \cite{Gao2008, HR2014}. Now we asymptotically expand the plasma dispersion function $ \mathcal{Z} (\zeta) = i \sigma \sqrt{\pi} \exp{( - \zeta^2)} - \zeta^{- 1} (1 + \zeta^{- 2}/2  + 3 \zeta^{- 4}/4  + 15 \zeta^{- 6}/8 + \cdots)$ as usual and neglect all terms of order higher than $\mathcal{O}(\zeta^{- 6})$, and then we arrive at the following simplified dispersion relation:
\begin{align}
\label{dis4}
\frac{1}{q^2} - \frac{\mathcal{G}_1}{\zeta^2} - \frac{\mathcal{G}_0}{\zeta^4} + i \sqrt{\pi} \chi \zeta^3 e^{- \zeta^2} = 0,
\end{align}
in which
\begin{align}
\mathcal{G}_1 = & \frac{2 + 4 \sigma + \sigma^2}{4} + \tau + \frac{(1 - \sigma)^2}{4 (1 + \tau)} + \frac{\tau}{2 q^2}\nonumber\\
 &\qquad + M^2 \bigg( 4 + \frac{\sigma - 1}{1 + \tau} \bigg) + \frac{M^4}{1 + \tau},\\
\mathcal{G}_0 = & \frac{15 + 8 \sigma}{8} + \frac{9 \tau}{8} + \frac{\sigma (\sigma - 1)}{4 (1 + \tau)} + \frac{3 \tau}{4 q^2} + \frac{\tau^2 (1 - \sigma^2)}{8 (1 + \tau)} \nonumber\\
& \qquad + M^2 \bigg(5 + \frac{\sigma - 1}{2 (1 + \tau)} \bigg) + \frac{M^4}{2 (1 + \tau)}.
\end{align}

By assuming $\zeta = q \Omega_K + i q \gamma_d$, where $\Omega_K$ and $\gamma_d$ are both real and are the normalized frequency and damping rate of GAM, respectively, Eq. \eqref{dis4} yields the frequency of GAM as
\begin{align}
\label{fre}
\Omega_K^2 = \frac{\mathcal{G}_1}{2} + \sqrt{\frac{\mathcal{G}_1^2}{4} + \frac{\mathcal{G}_0}{q^2} },
\end{align}
and the imaginary part of \eqref{dis4} gives the damping rate of GAM in a toroidally rotating anisotropic plasma
\begin{align}
\label{damp}
\gamma_d = - \frac{\sqrt{\pi} q^5 \Omega_K^6 }{2 \sqrt{\mathcal{G}_1^2 + 4 \mathcal{G}_0/q^2}} e^{- q^2 \Omega_K^2},
\end{align}
where $\chi = 1$ is adopted due to the factor that $ \gamma_d \ll \Omega_K$. The two equations above are the major results representing the effects of toroidal rotation on the GAM in anisotropic plasmas.

Eq. \eqref{fre} illustrates the dependence of GAM frequency on the safety factor $q$, temperature ratio $\tau$, anisotropy $\sigma$, and the Mach number $M$. For general Mach number $M \thicksim \mathcal{O}(1)$, by ignoring terms proportional to $1/q^2$, GAM frequency is reduced to
\begin{align}
\label{fre2}
\Omega_G^2 = \frac{1}{2} + \sigma & + \frac{\sigma^2}{4} + \tau + \frac{(1 - \sigma)^2}{4 (1 + \tau)} \nonumber\\
& + M^2 \bigg( 4 + \frac{\sigma - 1}{1 + \tau} \bigg) + \frac{M^4}{1 + \tau}.
\end{align}
According to the frequency above, we can see that the coefficient of $M^4$ depends only on $\tau$ and has no relation to anisotropy. The coefficient of $M^2$ relying on both $\sigma$ and $\tau$ is always positive and turns into $4$ in the isotropic case. Thereby, the toroidal rotation always increases the GAM frequency\cite{HR2015}.

Let us focus on the coupling effect of temperature ratio and the anisotropy by zeroing $M$. Eq. \eqref{fre2} is reduced to
\begin{align}
\label{fre3}
\Omega_{G 0}^2 = \frac{1}{2} + \sigma & + \frac{\sigma^2}{4} + \tau + \frac{(1 - \sigma)^2}{4 (1 + \tau)}.
\end{align}
In the isotropic limit with $\sigma = 1$, the frequency above reproduces the classical one $7/4 + \tau$. It is found that when $\sigma < 3 + 2 \tau$, there is $\partial \Omega_{G 0}^2/\partial \tau > 0$ and for $\sigma > 3 + 2 \tau$, $\partial \Omega_{G 0}^2/\partial \tau $ is negative. That is to say, when $\sigma < 3 + 2 \tau$, the increased electron temperature (with fixed $T_\parallel$) will increase the GAM frequency. When $\sigma$ is larger than $ 3 + 2 \tau$, which requires strong anisotropy since generally $\tau \geq 1$ in a realistic tokamak plasmas, the increasing of electron temperature decreases the GAM frequency. On the other hand, one can see $\partial \Omega_{G 0}/\partial \sigma$ is always positive, meaning that the anisotropy always tends to enlarge the GAM frequency. More stronger anisotropy means more greater $\delta p_\perp$ when $p_\parallel$ is fixed, leading to the increasing of the total perturbed pressure $\delta p_\parallel + \delta p_\perp$, which eventually determines the GAM frequency. Keeping terms to the order of $\mathcal{O}(1/q^2)$, Eq. \eqref{fre} yields
\begin{widetext}
\begin{align}
\Omega_G^2 = \Omega_{G 0}^2 \left[ 1 + \frac{30 + 16 \sigma + (22 + 8 \sigma + 2 \sigma^2) \tau + 8 \tau^2 + 2 (1 - \sigma) (\tau + \tau \sigma - 2 \sigma)}{q^2 \bigg( 2 + 4 \sigma + \sigma^2 + 4 \tau + \frac{(1 - \sigma)^2}{1 + \tau} \bigg)^2 } \right].
\end{align}
\end{widetext}
This equation illustrates the approximate frequency of GAM in anisotropic tokamak plasmas with arbitrary toroidal rotation in the large safety factor case.

Eq. \eqref{fre3} is not identical with the previous one reported in Ref. \onlinecite{Zhou2010}, which reads [see Eq. (23) in \onlinecite{Zhou2010}]
\begin{align}
\Omega_{G 1}^2 = \frac{3}{4} + \frac{\sigma}{2} + \frac{\sigma^2}{2} + \frac{1 + \sigma}{2} \tau.
\end{align}
Subtracted by the equation above, Eq. \eqref{fre3} leaves
\begin{align}
\Omega_{G0}^2 - \Omega_{G 1}^2 = \frac{(\sigma - 1) \tau (1 + \sigma + 2 \tau)}{4 (1 + \tau)},
\end{align}
which indicates that our result becomes the same as the on in Ref. \onlinecite{Zhou2010} only when $\tau = 0$ and/or $\sigma = 1$. Since Ref. \onlinecite{Zhou2010} used the standard bi-Maxwell distribution in which $T_\perp$ and $T_\parallel$ were both assumed to be the function of $\psi$, $\partial F_0/\partial \mu$ was neglected in the kinetic equation. However, a self-consistent distribution \eqref{fenbu} implies that $\frac{\partial F_0^i}{B \partial \mu} = (1 - \frac{1}{\sigma}) \frac{1}{T_\parallel} $, which can not be disregarded in the kinetic equation except for $\sigma \simeq 1$. It is also noted that $\partial F_0^i/\partial \mu$ only contributes to the first term on the right-hand side of Eq. \eqref{ionp}. In the case of $\tau = 0$, current neutrality condition $\langle J^r \rangle = 0$ is used to obtain the dispersion relation of GAM\cite{HR2014} and only the second term on the right-hand side of Eq. \eqref{ionp} plays a role. Hence the dependence of $F_0^i$ on the magnetic moment $\mu$ does not impact the final dispersion relation.

The damping rate \eqref{damp} is proportional to $e^{ - q^2 \Omega_K^2}$ and will be dramatically decreased by increasing $\Omega_K$. Considering that $\Omega_K$ always increased by $\sigma$ and decreases with $\tau$ when $ \tau < (\sigma - 3)/2$, we let $\sigma = 3$ and $\tau = 0$ to minimum $\Omega_K^2$, which is about $6.75$. As a result, the damping rate is nero zero and ignorable. For general anisotropy strength $\sigma \lesssim 3$, the temperature ratio $\tau$ is usually greater than $(\sigma - 3)/2$. Then the GAM frequency increases as $\tau$ grows up. It can be concluded that the collisionless damping rate is remarkably diminished by either $\tau$ or $\sigma$.

Geodesic acoustic mode (GAM) in a toroidally rotating anisotropic tokamak is investigated by using gyro-kinetic (GK) equation. Self-consistent equilibrium distribution function is derived according to the zeroth-order Fokker-Planck equation, leading to $T_\parallel = T_\parallel (\psi)$ and $T_\perp = T_\parallel B/(B - B_0(\psi))$ for ions and $T_e = T_e (\psi)$ for electrons, where $\tau \gg m_e/m_i$ is assumed. The perturbed distribution function is analytically obtained from the GK equation with $m = n = 0$ for the perturbed electrostatic potential $\delta \phi$. By focusing on the passing ions and assuming small drift orbit radius, the general dispersion relation of GAM is derived. For large safety factor, the GAM frequency is given as $\Omega_{G0}^2 = 1/2 + \sigma + \sigma^2/4 + (1 - \sigma)^2/[4 (1 + \tau)]$ to the leading order. The anisotropy strength $\sigma$ is shown to increase the GAM frequency and diminish the damping rate. The temperature ratio increases the GAM frequency when it is larger than $(\sigma - 3)/2$ and decreases the GAM frequency when $\tau < (\sigma - 3)/2$.

\begin{acknowledgments}
This work was supported by the China National Magnetic Confinement Fusion Science Program under Grant Nos. 2015GB120005 and 2013GB112011, and the National Natural Science Foundation of China No. 11275260.

\end{acknowledgments}


\begin{references}
\bibitem{book} R. D. Hazeltine and J. D. Meiss, \emph{Plasma Confinement} (Addison-Wesley, Redwood City, 1992).
\bibitem{Iter} A. Fasoli, C. Gormenzano, H. L. Berk, B. Breizman, S. Briguglio, D. S. Darrow, N. Gorelenkov, W. W. Heidbrink, A. Jaun, S. V. Konovalov, R. Nazikian, J.-M. Noterdaeme, S. Sharapov, K. Shinohara, D. Testa, K. Tobita, Y. Todo, G. Vlad and F. Zonca, Nucl. Fusion \textbf{47}, S264 (2007).
\bibitem{Jet1} G. A. Cottrell and D. F. H. Start, Nucl. Fusion \textbf{31}, 61 (1991).
\bibitem{Mast} M. J. Hole, G. von Nessi, M. Fitzgerald, K. G. McClements, J. Svensson and the MAST team, \ppcf \textbf{53}, 074201 (2011).
\bibitem{LHD} T. Yamaguchi, K. Y. Watanabe, S. Sakakibara, Y. Narushima, K. Narihara, T. Tokuzawa, K. Tanaka, I. Yamada, M. Osakabe, H. Yamada, K. Kawahata, K. Yamazaki and LHD Experimental Group, Nucl. Fusion \textbf{45}, L33 (2005).
\bibitem{1958} M. D. Kruskal and C. R. Oberman, Phys. Fluids \textbf{1}, 275 (1958).
\bibitem{1964} T. G. Northrop and K. J. Whiteman, \prl \textbf{12}, 639 (1964).
\bibitem{1967} H. Grad, Phys. Fluids \textbf{10}, 137 (1967).
\bibitem{1970} D. Dobrott and J. M. Greene, Phys. Fluids \textbf{13}, 2391 (1970).
\bibitem{1974} G. O. Spies and D. B. Nelson, Phys. Fluids \textbf{17}, 1865 (1974).
\bibitem{1974b} G. O. Spies and D. B. Nelson, Phys. Fluids \textbf{17}, 1879 (1974).
\bibitem{1978} D. B. Nelson, G. O. Spies, and C. L. Hedrick, Phys. Fluids \textbf{21}, 1742 (1978).
\bibitem{1980} W. A. Cooper, G. Bateman, D. B. Nelson, and T. Kammash, Nucl. Fusion \textbf{20}, 985 (1980).
\bibitem{1981} T. M. Antonsen Jr. and Y. C. Lee, Phys. Fluids \textbf{25}, 132 (1982).
\bibitem{1987} E. R. Salberta, R. C. Grimm, J. L. Johnson, J. Manickam, and W. M. Tang, Phys. Fluids \textbf{30}, 2796 (1987).
\bibitem{1990} R. Iacono, A. Bondeson, F. Troyon, and R. Gruber, Phys. Fluids B \textbf{2}, 1794 (1990).
\bibitem{2010} V. D. Pustovitov, \ppcf \textbf{52}, 065001 (2010).
\bibitem{2012} V. D. Pustovitov, AIP Conference Proceedings \textbf{1478}, 50 (2012).
\bibitem{2012b} M. J. Hole, G. von Nessi, M. Fitzgerald and the MAST team, \ppcf \textbf{55}, 014007 (2013).
\bibitem{2013} M. Fitzgerald, L. C. Appel, and M. J. Hole, Nucl. Fusion \textbf{53}, 113040 (2013).
\bibitem{2014} Z. S. Qu, M. Fitzgerald, and M. J. Hole, \ppcf \textbf{56}, 075007 (2014).
\bibitem{1976} J. W. Connor and R. J. Hastie, Phys. Fluids \textbf{19}, 1727 (1976).
\bibitem{1995} R. O. Dendy, R. J. Hastie, K. G. McClements, and T. J. Martin, \pop \textbf{2}, 1623 (1995).
\bibitem{Jet2} W. Zwingmann, L.-G. Eriksson, and P. Stubberfield, \ppcf \textbf{43}, 1441 (2001).
\bibitem{2003} E. V. Belova, N. N. Gorelenkov, and C. Z. Cheng, \pop \textbf{10}, 3240 (2003).
\bibitem{2004} L. Guazzotto, R. Betti, J. Manickam, and S. Kaye, \pop \textbf{11}, 604 (2004).
\bibitem{HR2014} H. Ren and J. Cao, \pop 21, 122512 (2014).

\bibitem{Diamond} P. H. Diamond, S.-I. Itoh, K. Itoh, and T. S. Hahm, \ppcf  \textbf{47}, R35 (2005).
\bibitem{winsor} N. Winsor, J. L. Johnson, and J. M. Dawson, Phys. Fluids \textbf{11}, 2448 (1968).


\bibitem{Zhou2010} M. Zhang and D. Zhou, Plasma Sci. Technol. \textbf{12}, 6 (2010).
\bibitem{Hinton1985} F. L. Hinton and S. K. Wong, Phys. Fluids \textbf{28}, 3082 (1985).
\bibitem{Chao} C. Dong, H. Ren, H. Cai, and D. Li, \pop 20, 102518 (2013).
\bibitem{GK} M. Artun M and W. M. Tang, \pop \textbf{1}, 2682 (1994).
\bibitem{CGK1968} P. H. Rutherford and E. A. Frieman, Phys. Fluids \textbf{11}, 569 (1968).
\bibitem{Gao2008} Z. Gao, K. Itoh, H. Sanuki, and J. Q. Dong, \pop \textbf{15}, 072511 (2008).
\bibitem{HR2015} H. Ren and J. Cao, arXiv:1501.01750.











%
%


\end{references}
\end{document}